\documentclass[10pt,conference]{IEEEtran}

\usepackage{amsmath,amsfonts}
\usepackage[utf8]{inputenc}
\usepackage{algorithm}
\usepackage{array}
\usepackage{algpseudocode}
\usepackage{todonotes}
\usepackage{graphicx}
\usepackage{svg}
\usepackage{textcomp}
\usepackage{svg}
\usepackage{xcolor}
\usepackage{longtable}
\usepackage{listings}
\usepackage{multirow}
\usepackage{lineno}

\def\BibTeX{{\rm B\kern-.05em{\sc i\kern-.025em b}\kern-.08em
    T\kern-.1667em\lower.7ex\hbox{E}\kern-.125emX}}
    
\usepackage{flushend}
\usepackage{tcolorbox}
\usepackage{color}
\usepackage{xspace}
\usepackage{tabularx}
\usepackage{booktabs}
\usepackage{longtable}
\usepackage[font=small,skip=0pt]{caption}
\usepackage{subcaption}
\usepackage{url}
\usepackage{tikz}
\usepackage{enumitem}
\usepackage{adjustbox}
\usepackage{stfloats}
\usepackage{tcolorbox}

\newcommand{\findingbox}[2]{\begin{tcolorbox}
[boxsep=3pt, left=4pt, right=4pt, top=2pt, bottom=2pt, colback=yellow!10, colframe=brown!40, coltext=black, title={#1}]
{#2}
\end{tcolorbox}}

\usepackage{hyperref}

\usepackage{makecell}
\usepackage{pifont}%
    
\newboolean{showcomments}
\setboolean{showcomments}{true}
\ifthenelse{\boolean{showcomments}}
 { \newcommand{\mynote}[2]{
      \fbox{\bfseries\sffamily\scriptsize#1}
        {\small$\blacktriangleright$\textsf{\emph{#2}}$\blacktriangleleft$}}}
        { \newcommand{\mynote}[2]{}}

\newcommand{\toolname}{{APAK}\xspace}

\definecolor{codegreen}{rgb}{0,0.6,0}
\definecolor{codegray}{rgb}{0.5,0.5,0.5}
\definecolor{codepurple}{rgb}{0.58,0,0.82}
\definecolor{backcolour}{rgb}{0.95,0.95,0.92}

\definecolor{lightgray}{rgb}{.9,.9,.9}
\definecolor{darkgray}{rgb}{.4,.4,.4}
\definecolor{purple}{rgb}{0.65, 0.12, 0.82}

\lstdefinelanguage{JavaScript}{
  keywords={typeof, new, true, false, catch, function, return, null, catch, switch, var, if, in, while, do, else, case, break},
  keywordstyle=\color{blue}\bfseries,
  ndkeywords={class, export, boolean, throw, implements, import, this},
  ndkeywordstyle=\color{darkgray}\bfseries,
  identifierstyle=\color{black},
  sensitive=false,
  comment=[l]{//},
  morecomment=[s]{/*}{*/},
  commentstyle=\color{purple}\ttfamily,
  stringstyle=\color{red}\ttfamily,
  morestring=[b]',
  morestring=[b]"
}

\usepackage{xspace}
\makeatletter
\DeclareRobustCommand\bmvaOneDot{\futurelet\@let@token\bmv@onedotaux}
\def\bmv@onedotaux{\ifx\@let@token.,\else.\null\fi\xspace}
\DeclareRobustCommand\bmvaTwoDot{\futurelet\@let@token\bmv@twodotaux}
\def\bmv@twodotaux{\ifx\@let@token.,\else.,\null\fi\xspace}
\makeatother

\begin{document}

\title{Context-Sensitive Pointer Analysis for ArkTS
}

\DeclareRobustCommand*{\IEEEauthorrefmark}[1]{%
    \raisebox{0pt}[0pt][0pt]{\textsuperscript{\footnotesize\ensuremath{#1}}}}

\author{
	\IEEEauthorblockN{
            Yizhuo Yang\IEEEauthorrefmark{1}, 
            Lingyun Xu\IEEEauthorrefmark{2}, 
            Mingyi Zhou\IEEEauthorrefmark{1}$^{\ast}$,
            Li Li\IEEEauthorrefmark{1}$^{\ast}$
        } 
        
	\IEEEauthorblockA{
        \IEEEauthorrefmark{1}
            \textit{School of Software, Beihang University}
    }
    \IEEEauthorblockA{
        \IEEEauthorrefmark{2}
            \textit{CBG Software Engineering Department, Huawei}
    }
}

\maketitle
\footnotetext[0]{$^{\ast}$Corresponding authors}
\begin{abstract}
Current call graph generation methods for ArkTS, a new programming language for OpenHarmony, exhibit precision limitations when supporting advanced static analysis tasks such as data flow analysis and vulnerability pattern detection, while the workflow of traditional JavaScript(JS)/TypeScript(TS) analysis tools fails to interpret ArkUI component tree semantics. The core technical bottleneck originates from the closure mechanisms inherent in TypeScript's dynamic language features and the interaction patterns involving OpenHarmony's framework APIs. Existing static analysis tools for ArkTS struggle to achieve effective tracking and precise deduction of object reference relationships, leading to topological fractures in call graph reachability and diminished analysis coverage. This technical limitation fundamentally constrains the implementation of advanced program analysis techniques.

Therefore, in this paper, we propose a tool named ArkAnalyzer Pointer Analysis Kit (\toolname), the first context-sensitive pointer analysis framework specifically designed for ArkTS. 
\toolname addresses these challenges through a unique ArkTS heap object model and a highly extensible plugin architecture, ensuring future adaptability to the evolving OpenHarmony ecosystem.
In the evaluation, we construct a dataset from 1,663 real-world applications in the OpenHarmony ecosystem to evaluate \toolname, demonstrating \toolname's superior performance over CHA/RTA approaches in critical metrics including valid edge coverage (e.g., a 7.1\% reduction compared to CHA and a 34.2\% increase over RTA).
The improvement in edge coverage systematically reduces false positive rates from 20\% to 2\%, enabling future exploration of establishing more complex program analysis tools based on our framework. Our proposed \toolname has been merged into the official static analysis framework ArkAnalyzer for OpenHarmony.


\end{abstract}

\section{Introduction}

The rapid development of OpenHarmony~\cite{openharmony_project, Li2025software}, an open-source, all-scenario operating system, has fostered a diverse application ecosystem. As applications grow in complexity, ensuring their correctness and security has become progressively challenging~\cite{Chen2024gui, Ma2024cid4hmos}. 
Static program analysis identifies potential issues without execution. 
Pointer analysis~\cite{andersen1994program}, central to many of these approaches, plays a vital role in understanding pointer usage within an application~\cite{Rupta2024}, aiding in the detection of bugs~\cite{Rahaman2019CryptoGuard}, memory leaks and vulnerabilities such as buffer overflows. Nevertheless, due to the inherent complexity of modern applications, especially OpenHarmony's integration of heterogeneous components and services, effective pointer analysis remains a non-trivial task. This paper explores pointer analysis challenges and methodologies for OpenHarmony static analysis.

However, existing static analysis tools for OpenHarmony focus solely on the static information from statements, leading to analysis accuracy issues. When the code contains statements that require dynamic execution for their full resolution, issues such as false positives or false negatives in call graph construction and incomplete type inference may arise. These issues adversely affect downstream OpenHarmony static analysis tools, like data flow analysis, which depend on accurate input.

Therefore, implementing pointer analysis within the existing OpenHarmony static analysis framework is urgently needed~\cite{li2017static}. However, existing pointer analysis methods (e.g., Soot~\cite{vallee2010soot} and Doop~\cite{bravenboer2009strictly, Doop2025} for Java and SVF~\cite{Sui2012valueflowanalysis, Sui2016SVF} for C/C++) are not compatible with the ArkTS program.
This difficulty stems from ArkTS's unique features, syntactic sugar, and the distinct architecture of the OpenHarmony framework. In addition, ArkTS introduces a unique declarative UI paradigm. OpenHarmony kernel's rich APIs present significant data flow modeling and tracking challenges. 


To address these challenges, this paper introduces \toolname, the first context-sensitive Andersen-style~\cite{andersen1994program} pointer analysis framework. It is designed to tackle accuracy issues within ArkAnalyzer~\cite{chen2025arkanalyzer}, an existing static analysis framework for OpenHarmony. 
A core design principle of \toolname is its extensibility, realized through a modular plugin system. This architecture is critical for future-proofing the analysis against the rapid evolution of the OpenHarmony ecosystem, allowing new framework APIs and language constructs to be supported without altering the core engine. 
For OpenHarmony APIs, \toolname adopts a specialized heap abstraction method. This method treats instantiated classes, function pointers, containers, and other relevant elements as heap objects~\cite{Kanvar2017heapabstraction}, simplifying pointer passing and invocation. Furthermore, \toolname differentiates between dynamic and static call statements to apply distinct context strategies. 
\toolname applies different context strategies for various scenarios and constructing an on-the-fly call graph with callsite-sensitivity~\cite{pnueli1981two, Might2010k-CFA-paradox} and function-sensitivity. 
This approach provides more accurate type and call graph information, particularly when statically analyzing dynamic language features. 
\toolname has been integrated into the official ArkAnalyzer open-source project, providing more precise type and call graph information for OpenHarmony applications and supporting security analysis.

To validate APAK's effectiveness in call graph generation, we constructed a test set of 1,663 real-world, open-source OpenHarmony apps and conducted comparative experiments with the Class Hierarchy Analysis (CHA) and Rapid Type Analysis (RTA) algorithms~\cite{Bacon1996CHARTA} integrated within the ArkAnalyzer. The results indicate \toolname achieves call edge set sizes comparable to CHA while demonstrating a 34.2\% improvement over RTA, all while maintaining efficient analytical performance.
It also demonstrates call resolution accuracy improvements of 5.1\% to 49.8\% across 12 samples and reduces the false positive rate from 20\% to 2\%. 
These findings substantiate the practicality of the APAK framework in analyzing real-world OpenHarmony apps, and establish reliable foundational support for future program analysis tools.


The main contributions we made are summarized as follows:
\begin{enumerate}[leftmargin=*]
    \item We proposed the first context-sensitive pointer analysis framework \toolname that is designed for ArkTS.
    \item Our method customizes heap abstraction and context-sensitive strategies for OpenHarmony applications, addressing the need for accurate type inference and call graph construction.
    \item We evaluated the proposed \toolname, demonstrating its effectiveness, efficiency, and utility in deriving accurate types and generating precise call graphs in OpenHarmony apps. We open-sourced the collected real-world OpenHarmony apps to advance future research\footnote{The dataset is available at \url{https://zenodo.org/records/15025243}}.
    \item We open-sourced our proposed \toolname by merging it into the official static analysis framework ArkAnalyzer for OpenHarmony\footnote{The repository is available at \url{https://anonymous.4open.science/r/APAK-1E45}}.
\end{enumerate}

\section{Background}

\subsection{ArkTS and TS}

\begin{figure}
    \centering
    \includegraphics[width=1\linewidth]{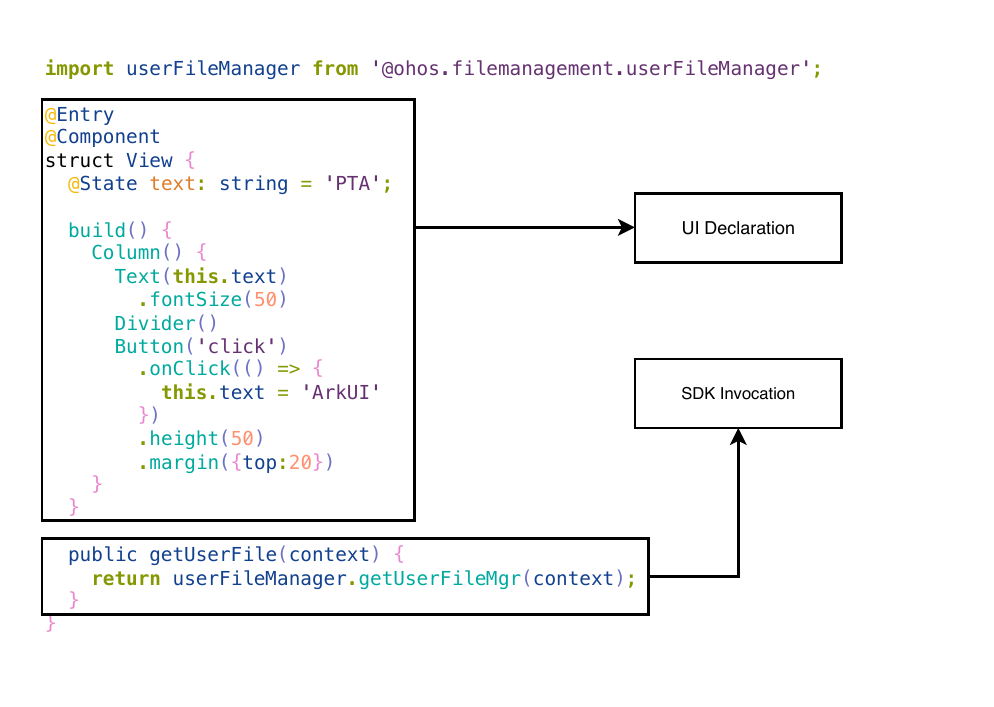}
    \caption{ArkTS code example.}
    \label{fig:arkts_code_example}
    \vspace{-2em}
\end{figure}

As a superset of TypeScript, ArkTS maintains core language characteristics while implementing systematic adaptations for mobile high-performance scenarios. Key architectural evolutions manifest in: 1) Paradigm innovation of declarative UI framework ArkUI. 2) System-level extension support through OpenHarmony public SDK. Figure~\ref{fig:arkts_code_example} shows an example of the unique features of ArkTS.

\noindent \textbf{Key Differences:} (1) ArkUI employs declarative syntax structures for UI description logic. By abstracting underlying rendering mechanisms (Virtual DOM diff algorithms, GPU instruction pipeline scheduling, etc.), it enables developers to focus on business logic construction. The framework automatically synchronizes interface state changes to rendering pipelines through DSL-layer semantic mapping. 
In the ArkUI framework, component declaration is achieved through the \texttt{struct} keyword, working in conjunction with decorators such as \texttt{@Entry} and \texttt{@Component} to complete component type annotation and enable diversified component definitions including entry components and custom components.

(2) The OpenHarmony Public SDK provides cross-device collaboration API collections through layered architecture design (including device abstraction layer, distributed capability middleware, atomic service interfaces~\cite{hmsos}). Its standardized interface contracts support "develop once, deploy everywhere" paradigm, significantly reducing multi-device adaptation complexity. The OpenHarmony SDK establishes a runtime support infrastructure for componentized architectures through the definition of core component libraries (e.g., \texttt{Column}, \texttt{Text}), while creating cross-layer system service access channels. The \texttt{userFileManager} interface adopts an abstraction layer design for persistent data storage, providing developers with a programmable interface paradigm at the application layer.

\subsection{Call Graph}
The call graph, as a core data structure in program analysis, formally characterizes invocation relationships between program units (nodes as entities, edges as potential flows).
It is fundamental for code optimization~\cite{chambers1996whole}, security defect detection~\cite{Reps1995graphreachability, Minami2023malware}, and architectural maintainability analysis~\cite{Feldthaus2013Efficient}. For compiler-driven optimizations, call graphs enable precise interprocedural analysis to guide decisions on function inlining, dead code elimination, and parallelization strategies. In security contexts, they underpin taint analysis by tracing vulnerability propagation paths across method boundaries. This is particularly crucial for identifying critical attack surfaces in Android applications, where incomplete call graph construction reportedly renders 58\% of apps unanalyzable~\cite{Samhi2024CallGraphSoundness}. 
In this study, we will focus on leveraging pointer analysis to enhance the accuracy of call graph construction within OpenHarmony.

\subsection{Pointer Analysis}

Pointer analysis (or points-to analysis)~\cite{sridharan2013alias, Weihl1980Interprocedural, Yannis2015Foundations} is a foundational static analysis technique that systematically determines the set of memory objects~\cite{Kanvar2017heapabstraction, tan2017} a pointer variable or expression may reference during program execution. 
It models relationships between pointers and their targets in memory, often by constructing a pointer assignment graph (PAG) where nodes are memory locations and edges denote possible references~\cite{Pereira2009wave}.

Modern pointer analysis tools exhibit distinct architectural characteristics and optimization strategies tailored for specific programming paradigms.
SVF specializes in C program analysis by leveraging the LLVM intermediate representation to perform scalable and precise interprocedural value-flow analysis. It constructs value flow graphs~\cite{Sui2012valueflowanalysis} for security vulnerability detection utilizing optimizations such as sparse value flow graph construction to reduce memory consumption. Tai-e~\cite{tan2023tai}, a developer-friendly framework designed for Java ecosystems, achieves higher precision in call graph construction compared to traditional tools. Doop~\cite{PointerAnalysis2015yannis} leverages Datalog-based context-sensitive analysis for Android application security vetting, while Rupta~\cite{Rupta2024} implements pointer analysis through Rust MIR for call graph construction.

\section{Motivating Example}

\begin{figure*}[!t]
    \centering
    \includegraphics[width=1\linewidth]{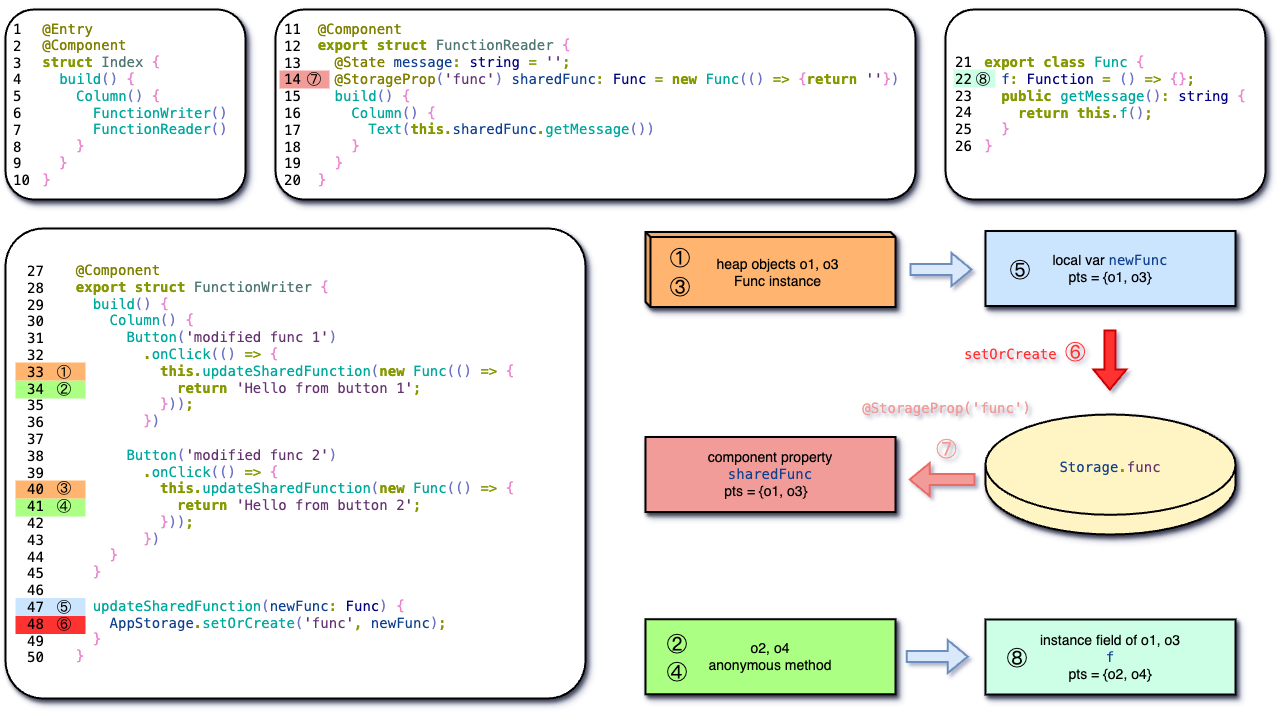}
    \caption{ArkTS code with AppStorage and Conceptual PAG.}
    \label{fig:motivating}
\end{figure*}

Constructing precise call graphs for ArkTS applications is challenging when they leverage centralized reactive data stores for state management and intercomponent data synchronization. Common in modern UI frameworks, these stores allow disparate application parts to share and react to state changes without direct dependencies or explicit instance passing. Components typically write complex data, including object instances encapsulating functional logic, to a globally accessible location using a key. Other components can then subscribe to this key and reactively receive updates whenever the associated data changes.

As demonstrated in Figure~\ref{fig:motivating}, an architectural pattern is implemented using \texttt{AppStorage}, which provides a key-value mechanism at the system level for state management and synchronization throughout the application. When a button in the \texttt{FunctionWriter} component is activated, its \texttt{updateSharedFunction} method dynamically creates a new \texttt{Func} object instance, which contains a function property. This instance is then published to this centralized reactive store under the key \texttt{func}. Correspondingly, the \texttt{FunctionReader} component declares a reactive link to the \texttt{func} key in the store. This link ensures that \texttt{FunctionReader}'s local \texttt{sharedFunc} property is automatically updated to reference the latest \texttt{Func} instance dispatched by \texttt{FunctionWriter}. The displayed text is then derived from an invocation of \texttt{this.sharedFunc.getMessage()}, which executes the \texttt{this.f()} function property of the currently shared \texttt{Func} instance, reflecting the state managed through the centralized store.

Constructing a precise call graph for the scenario in Figure~\ref{fig:motivating} presents a considerable challenge for existing static analysis tools within the OpenHarmony ecosystem. Conventional methods like CHA or RTA, the baseline algorithms in ArkAnalyzer, have fundamental limitations here. These methods typically struggle to accurately resolve the concrete object instance that the \texttt{sharedFunc} variable in \texttt{FunctionReader} points to when \texttt{getMessage()} is invoked. Because the instance of \texttt{Func} object is retrieved dynamically from the application's \texttt{AppStorage} (which is updated by button clicks in \texttt{FunctionWriter}), CHA/RTA lack the necessary pointer or data flow resolution capabilities to precisely track which specific \texttt{Func} instance (and therefore which encapsulated \texttt{f} function closure) is active. Consequently, they cannot reliably determine the precise target of the \texttt{this.f()} call within the \texttt{Func.getMessage()} method, leading to an imprecise or incomplete call graph where the true callees (the dynamically defined closures) are either missed or over-approximated.

\ding{172}, \ding{174} are \texttt{Func} object instances created in \texttt{FunctionWriter}. APAK abstracts them as distinct heap objects (denoted as $o_1$ and $o_3$).

\ding{173}, \ding{175} are lambda expressions in lines 33 and 40. They are also abstracted as heap objects (denoted as $o_2$ and $o_4$), and then passed to these \texttt{Func} constructors. These anonymous function objects become  targets of the \texttt{f} field within their respective \texttt{Func} instances. APAK establishes that the points-to set of $o_1.f$ is {$o_2$} (i.e., $pts(o_1.f) = \{o_2\}$), and similarly, $pts(o_3.f) = \{o_4\}$.

\ding{176} represents the parameter \texttt{newFunc} in the method \texttt{updateSharedFunction}. When \texttt{updateSharedFunction} is called, the parameter receives a reference to one of these \texttt{Func} instances. $pts(newFunc)$ can be considered as potentially containing \{$o_1, o_3$\} with different potential executions initiated by button clicks.

\ding{177} represents the crucial step that occurs with \texttt{AppStorage.setOrCreate}. APAK models the \texttt{func} key within \texttt{AppStorage} as a specific storage location. The pointer analysis rule for \texttt{setOrCreate} propagates the points-to set of \texttt{newFunc} to this storage location. Thus, $pts(AppStorage.func)$ becomes \{$o_1, o_3$\}

\ding{178} represents the \texttt{@StorageProp('func')} decorator on the \texttt{sharedFunc} property in the \texttt{FunctionReader} component, meaning the variable is synchronized with \texttt{AppStorage.func}. APAK establishes a pointer-flow relationship from \texttt{AppStorage.func} to \texttt{sharedFunc}. This means $pts(sharedFunc)$ will also be \{$o_1, o_3$\}.

This precise tracking of pointer propagation through heap abstractions for class instances and functions, along with specialized modeling of \texttt{AppStorage} interactions, allows APAK to identify the correct potential callees ($o_2$ or $o_4$) for the indirect function call, thus constructing an accurate call graph where traditional methods would falter.

In essence, the motivating example underscores the core difficulties for static analysis in ArkTS: precisely tracking dynamically created objects and their encapsulated logic (like closures) as they propagate indirectly through framework-specific reactive data stores such as AppStorage, and subsequently, accurately resolving indirect call targets that depend on these dynamically managed objects.

\section{APAK: Pointer Analysis for ArkTS}
\subsection{Overview}
To address the technical challenges discussed in previous sections, we propose \toolname – the first pointer analysis framework for ArkTS. Grounded in the formal semantics of the ArkTS language specifications, \toolname employs a constraint-driven analysis algorithm augmented with incremental rule extensions to achieve precise modeling of OpenHarmony application characteristics.

\toolname's core innovation lies in 
(1) a highly extensible plugin architecture: \toolname's design allows for the integration of custom, rule-based plugins to analyze new or specialized APIs, ensuring future-proof adaptability.
(2) semantical model of the features of ArkTS (e.g., the store mechanism), which allows \toolname to understand and precisely model the behavior of OpenHarmony applications.
This chapter will sequentially present the language model of ArkTS programs (Section \ref{sec:arkts_language_model}), the core algorithm for pointer propagation (Section \ref{sec:propagation_algorithm}), the extension mechanism of the analysis algorithm (Section \ref{sec:extension_arkts}), and the context-sensitive processing mechanism (Section \ref{sec:context_sensitive}).

\subsection{ArkTS Language Model}
\label{sec:arkts_language_model}

\toolname adopts ArkIR, an intermediate representation (IR) of ArkTS source code implemented within ArkAnalyzer~\cite{chen2025arkanalyzer}, as its input. Serving as an intermediate abstraction layer, ArkIR performs semantic normalization of source code through decoupling and desugaring processes. Table~\ref{tab:arkts_language_model} illustrates a simplified syntactic specification of ArkIR along with the formal description of its type system. The model intentionally excludes the control flow statements (e.g., conditional branches, loop structures) to focus on core semantic features for pointer analysis.

\begin{table}
  \centering
  \caption{Simplified language model of ArkTS.}
  \begin{minipage}{\columnwidth}
  \begin{tabular}{@{\hspace{2em}}lll@{}}
       $t \in$ Type   & ::= & \texttt{number} $\mid$ \texttt{string} $\mid$ \texttt{boolean} $\mid$ \texttt{T} \\
       & &$\mid$ \texttt{(t|t)} $\mid$ ... \\
       $e \in$ Expr   & ::= & \texttt{e.x} $\mid$ \texttt{new x(\{e\})} $\mid$ \texttt{(\{e\}) => e} \\
       $s \in$ Stmt   & ::= & \texttt{let x: t = e} $\mid$ \texttt{const x = e} \\                  &     & $\mid$ \texttt{e.method(\{e\})} $\mid$ \texttt{x(\{e\})} \\ & & $\mid$ \texttt{e.x := e}
       $\mid$ \texttt{x := e} $\mid$ \texttt{return e} \\
       $f \in$ Function & ::= & \texttt{function x(\{e\}): t \{\{s\}\}} \\
       $c \in$ Class  & ::= & \texttt{\{d\} class x(\{e\}) \{\{m\}\}} \\
       $st \in$ Struct & ::= & \texttt{\{d\} struct x(\{e\}) \{\{m\}\}} \\
       $m \in$ Method & ::= & \texttt{method x(\{e\}): t \{\{s\}\}} \\
       $n \in$ Namespace & ::= & $m$ $\mid$ \texttt{\{s | f | c\}} \\
       $d \in$ Decorator & ::= & \texttt{@Component} $\mid$ \texttt{@Entry} $\mid$ ... \\
  \end{tabular}
  \label{tab:arkts_language_model}
  \end{minipage}
  \vspace{-2em}
\end{table}

The language model adopts a hierarchical architecture organized as follows:
\begin{itemize}[leftmargin=*]
    \item Declaration Layer: Includes declaration units of namespaces, functions, classes, and their methods.
    \item Statement Layer: Composes basic execution units, with focused modeling on statement types critical to pointer analysis: \texttt{AssignmentStmt} (Drives propagation of pointer aliasing relationships), \texttt{PropertyAccessStmt} (Involves pointer dereferencing on object fields) and \texttt{CallStmt} (Handling interprocedural propagation).
    \item Expression Layer: At the statement granularity level, expressions are decomposed into atomic operational units, \texttt{NewExpr} and \texttt{LambdaExpr} serve as the core semantics for heap object allocation.
\end{itemize}

ArkTS builds upon TypeScript’s structural typing by introducing mandatory type declarations, which explicitly constrain the type boundaries for variables and parameters, particularly those used as pointers. This enhances the foundation for type inference in static analysis. In addition, ArkTS extends support for ArkUI features, such as struct components and decorators. Struct components serve as state containers for UI elements, and their memory layouts have a significant impact on pointer reachability analysis. Meanwhile, decorators, which inject lifecycle hooks through metaprogramming, necessitate special handling of their implicit pointer operations.

The language model demonstrates comprehensive coverage of ArkIR while effectively filtering out non-pointer-related statements. Building upon this foundation, we constructed the core rule set for interprocedural pointer propagation.

\subsection{Propagation Algorithm}
\label{sec:propagation_algorithm}

\toolname models set constraints of the ArkIR through a PAG. Its symbolic notations are defined in Table~\ref{tab:pointer_analysis_notation}. The pointer domain $P$ in the program comprises both program variables and object field pointers, formally defined as $P = V \bigcup (O \times F)$. $V$ contains all program variables (including local variables, global variables, and formal parameters). $O$ denotes the set of runtime-allocated heap objects, and $F$ represents the set of object fields. Each pointer $p \in P$ is associated with a points-to set $pts(p)$, which statically records all abstract heap objects potentially referenced by pointer $p$ during program execution, whereby any object $o \in O$ becomes reachable through pointer $p$.

The graph structure contains two node categories: (1) program pointer nodes (including variables, formal parameters, etc.) whose points-to sets characterize potential heap object references, and (2) heap object nodes generated via heap abstraction, visually distinguished using double-circle notation. Directed edges represent inclusion constraints between pointers, formalized as $\forall(p_i, p_j) \in E$, $pts(p_i) \subseteq pts(p_j)$ to ensure semantic correctness in value propagation.

\begin{table}
    \centering
    \caption{Pointer analysis notation system and domain definitions. We use these symbols to formalize the propagation rules introduced in this paper.}
    \begin{tabular}{ccc} 
        \hline 
        Syntactic Element & Notation & Domain\\ 
        \hline 
        local & $x$, $y$ & $V$\\ 
        object & $o_i$, $o_j$ & $O$\\ 
        method & $m$ & $M$\\ 
        function & $func$ & $Func$\\ 
        field & $f$ & $F$\\ 
        pointer & $p$ & $P$\\
        pointer set & $pts$ & $\mathcal{P}(O)$\\ 
        \hline
    \end{tabular}
    \label{tab:pointer_analysis_notation}
    \vspace{-2em}
\end{table}

The edge generation process is implemented through semantic parsing of ArkIR instructions. ArkIR opcode classifier categorizes program statements into eight distinct pointer operation patterns, followed by the propagation rule set defined in Table~\ref{tab:propagation_rule} to generate corresponding constraint edges under different context environment (denoted as $c$). Through iterative solving of the constraint system's reachability properties, the pointer analysis converges to a fixed-point solution, ultimately establishing the complete PAG. Building upon this, Section \ref{sec:extension_arkts} will present a plugin framework of ArkTS's advanced language features. Section \ref{sec:context_sensitive} will elaborate on hybrid context-sensitivity strategies.

\begin{table*}
    \centering
    \caption{Basic Pointer analysis propagation rules.}
    \begin{adjustbox}{max width=\textwidth, center}
    \renewcommand{\arraystretch}{1.5}
    \begin{tabular}{ccccc} \cline{1-5}
         Kind&  Operation&  Statement&  Propagation Rule& PAG Edge \\ \cline{1-5}
         
         \multirow{2}{*}{alloc} & create object&  i: \texttt{let v = new T()}& \multirow{2}{*}{$c:o_i \in pts(c:v)$} & \\
         
         & create function pointer &  i: \texttt{let v = () => \{\}}&  & \\ \cline{1-5}
         
         assign &  variable assign & y = x & $\begin{gathered}
            c:o_i \in pts(c:x) \xrightarrow{} c':o_i \in pts(c':y)
         \end{gathered}$ & $x \xrightarrow{} y$ \\ \cline{1-5}
         
         store &  field store & \texttt{y.f = x} & $\begin{gathered}
            c:o_i \in pts(c:x), c':o_j \in pts(c':y) \xrightarrow{} c'':o_i \in pts(c'':o_j.f)
         \end{gathered}$ & $x \xrightarrow{} o_j.f$ \\
         
         load &  field load & \texttt{x = y.f} & 
         $\begin{gathered}
            c:o_i \in pts(c:y), c':o_j \in pts(c':o_i.f) \xrightarrow{} c'':o_j \in pts(c'':x)
         \end{gathered}$ & $o_i.f \xrightarrow{} x$ \\ \cline{1-5}
         
         \multirow{3}{*}{call} & static call & \texttt{y = func(v)} & 
         \multirow{3}{*}{
             $\begin{gathered}
                 f = Dispatch(), c' = [f::c] \\
                 c:o_i \in pts(c:v), c':o_j \in pts(c':v_{return}) \xrightarrow{} \\
                 c'':o_i \in pts(c'':v_{param}), c''':o_j \in pts(c''':y)
             \end{gathered}$
         } &
         \multirow{3}{*}{
             $\begin{gathered}
                v \xrightarrow{} v_{param} \\
                v_{return} \xrightarrow{} y
             \end{gathered}$
         } \\
         
         & dynamic call & \texttt{y = x.method(v)} &  & \\
         
         & function pointer call & \texttt{y = p()} &  & \\ \cline{1-5}
         
    \end{tabular}
    \end{adjustbox}
    \label{tab:propagation_rule}
    \vspace{-1em}
\end{table*}

The alloc rule uniformly handles memory allocation operations in programs. It covers two scenarios: object instantiation and function pointer management. For \texttt{new} expressions involving classes/interfaces, the type inference module extracts the explicit type $T$ from the statement and generates a corresponding heap object $o_i$ ($i$ is the code line number) . \toolname establishes the reference relationship between variables and heap objects through the assignment constraint $o_i \xrightarrow{} v$. In the case of lambda expressions, ArkIR represents them as anonymous method variables (e.g., \texttt{anonymous\_method\_1}). The type system annotates these variables with \texttt{FunctionType} while creating special function objects $o_i$ during heap abstraction. When parsing assignments in the form of \texttt{let v = anonymous\_method\_1}, it not only establishes the pointing constraint $o_i \xrightarrow{} v$ but also records the corresponding IR definition node in $o_i$. This implementation enables bidirectional data flow tracking between lambda expression declarations and their call targets. Such dual-processing mechanisms ensure static analysis can accurately capture memory behavior characteristics under both object-oriented and functional programming paradigms.

The assign rule, the core mechanism for resolving assignment semantics, establishes dataflow edges within the PAG. When processing assignment statements of the form \texttt{y = x}, this rule establishes a directed dataflow edge from the right side pointer $p_x$ to the left side pointer $p_y$ (denoted as $p_x \xrightarrow{} p_y$) within the PAG. Concurrently, it enforces pointer set propagation through the set inclusion relationship $pts(x) \subseteq pts(y)$. The load/store rules govern the dynamic resolution of object field access operations. When processing field access statements of the form \texttt{x = y.f}, the rules first retrieve the $pts(y)$ of the base pointer $y$ through traversal of the PAG. For each heap object instance $o_i \in pts(y)$, it dynamically generates a corresponding field-sensitive pointer $o_i.f$. This process ultimately constructs dataflow propagation paths from the field-sensitive pointers $o_i.f$ to the left-hand side variable $x$.

The call rule adopts a phased analysis strategy that employs distinct resolution strategies for three invocation types mentioned before. Static invocations are resolved through direct symbol resolution to determine the target method body. Dynamic invocations \texttt{o.m()} dynamically compute the $pts(o)$ of receiver objects during analysis iterations, subsequently resolving concrete instance methods via virtual method table (VMT)\cite{Driesen1996VMT} lookups. Function pointer invocations require inspection of points-to set, where anonymous functions are retrieved from contained \texttt{FunctionType} objects. Algorithm \ref{alg:propagation_algorithm} demonstrates how the pointer analysis framework handles these call types through stage-specific resolution logic in \toolname.

\begin{algorithm}[t]
\caption{Propagation algorithm} \label{alg:propagation_algorithm}
\begin{algorithmic}[1]
\Require 
  \Statex \texttt{entries} $\gets$ ability and component lifecycle methods collected from project

\Ensure
  \Statex $\triangleright$ PAG
  \Statex $\triangleright$ CG
  
\Procedure{start}{workList $= [entries]$}
    \While{workList not empty}
        \State \Call{initWorkItem}{} 
        \State \Call{solveConstraints}{}
        \State \Call{solveDynamicCall}{}
        \State \Call{solveFunctionPointerCall}{}
    \EndWhile
\EndProcedure
\end{algorithmic}
\end{algorithm}

The pointer analysis propagation algorithm implements an iterative analysis framework, which is organized based on method-level processing units. During initialization, ArkAnalyzer collects lifecycle methods of all abilities and components in \texttt{DummyMain}~\cite{Arzt2014flowdroid} as the project entry method. The algorithm enqueues the project's entry method into the \texttt{worklist} as the initial processing target. The analysis subsequently enters a multi-iteration processing loop. For the method $m$ in the worklist:

\begin{enumerate}[leftmargin=*]
    \item \texttt{initWorkItem}: Generates method-local PAG while deferring processing of dynamic method invocations and function pointer invocations and storing them in a pending call queue. Static calls are immediately resolved through symbol resolution to establish interprocedural edges.
    \item \texttt{solveConstraints}: Iterates over assignment and load/store statements within the method. It applies pointer propagation rules to transfer and merge points-to sets. The monotonicity of set inclusion relationships guarantees rapid convergence to local fixed-point solutions.
    \item \texttt{solveDynamicCall}, \texttt{solveFunctionPointerCall}: For dynamic method invocations, retrieves VMT through receiver objects' points-to sets to derive concrete target methods. For function pointer invocations, matches executable functions via type signature mapping tables under current context constraints.
\end{enumerate}

Newly resolved methods from call sites are injected back into the \texttt{workList}, driving iterative advancement of cross-procedural analysis. When the worklist becomes empty, the iteration terminates and returns the completed PAG and CG.

\subsection{Extensibility for ArkTS Language Features}
\label{sec:extension_arkts}

\begin{figure}
    \centering
    \includegraphics[width=1\linewidth]{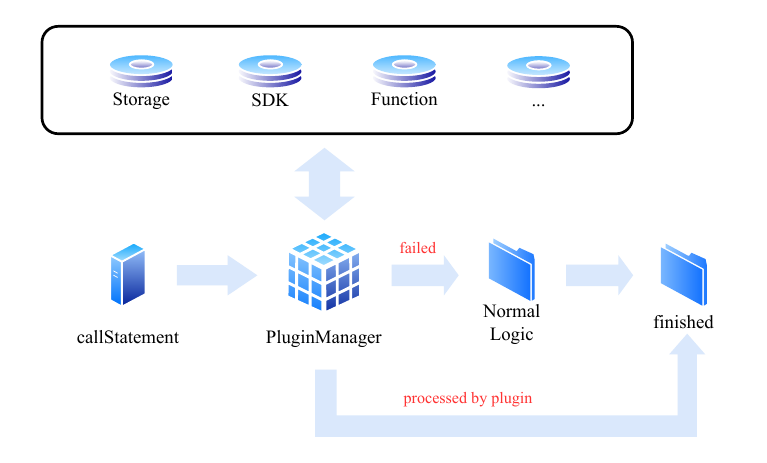}
    \caption{Plugin System Framework.}
    \label{fig:plugin_framework}
    \vspace{-1.4em}
\end{figure}

A core design principle of APAK is its extensibility, which is achieved through a highly modular plugin system in Figure~\ref{fig:plugin_framework}. This architecture is crucial to keeping up with the rapidly evolving OpenHarmony ecosystem. Instead of a monolithic design, \toolname allows new language features and framework APIs to be supported by adding new plugins, without altering the core analysis engine.

The plugin system is designed to handle API calls to both external libraries (e.g., the OpenHarmony SDK) and TypeScript's built-in libraries. 
The architecture is illustrated in Figure~\ref{fig:plugin_framework}. 
During the \texttt{solveDynamicCall} and \texttt{solveFunctionPointerCall} stage, the plugin manager extracts the signature of a call statement and checks for a registered plugin capable of handling that API type. 
The corresponding plugin then invokes the PAG and CG interfaces according to its specific rules. 
Finally, it returns the result to the manager, allowing the analysis to proceed to the subsequent statements.

\subsubsection{AppStorage \& LocalStorage Plugin}

\texttt{AppStorage} and \texttt{LocalStorage} are built-in storage mechanisms in ArkTS, offering cross-component data access and state synchronization, albeit with different scopes. 
\texttt{AppStorage} provides an application-level singleton for UI state, while \texttt{LocalStorage} is typically scoped more locally (e.g., to a specific component or custom class instance). 
Despite these scope differences, both employ a similar set of APIs and underlying principles for property storage, as exemplified by \texttt{AppStorage} in Listing~\ref{code:appstorage}. 
Developers can use these APIs (like \texttt{setOrCreate}, \texttt{get}, \texttt{Link}, and \texttt{Prop}, which are common to both) to register variable instances, function pointers, and other objects into their respective storage contexts. 
The combination of direct writability to these storage objects and the propagation of references they hold introduces significant challenges for pointer analysis.

\begin{lstlisting}[caption={AppStorage Usage.}, label=code:appstorage]
    function set() {
        AppStorage.setOrCreate(x, 1);
    }
    function getOrSynchronize() {
        let x = AppStorage.get(x); // x = 1
        let link1 = AppStorage.Link('x');
        let link2 = AppStorage.Link('x');
        let prop = AppStorage.Prop('x');

        link1.set(2);   // two-way sync: link1=link2=prop=2
        prop.set(3);    // one-way sync: link1=link2=2, prop=3
        link2.set(4);   // two-way sync: link1=link2=prop=4
    }
\end{lstlisting}

The \textit{StoragePlugin} employs signature matching to identify storage-related API invocation patterns, activating distinct processing strategies. 
For storage creation operations, it generates field nodes with isolated storage domains, whose points-to sets maintain the references to initialized objects. 
To implement differentiated synchronization between \texttt{prop} and \texttt{link} mechanisms, the system innovatively deploys backflow edge injection. 
This strategy involves selectively adding dataflow edges to the PAG that run in the opposite direction of typical value propagation (i.e., from a local variable back to a storage location) to model specific data binding semantics, as detailed for \texttt{Link} and \texttt{Prop} below.

For \texttt{link} synchronization, \toolname establishes bidirectional edges between parameter variable \texttt{p} and \texttt{AppStorage.x}, forming Strongly Connected Components (SCC) in the PAG (denoted as $SCC(N, E)$, $N$ represents node set and $E$ edge set). 
This ensures mutual reference synchronization.
For \texttt{prop} synchronization, the base strategy is retained. 
Without backflow edges in the topology, its points-to set only supports unidirectional data flow, preventing local modifications from propagating back to source attribute nodes.

\subsubsection{OpenHarmony SDK Invocation Plugin}

The OpenHarmony Software Development Kit (SDK)\cite{hmsos}, as a standardized capability collection for OpenHarmony native applications and meta-service development, provides modular development interfaces covering application frameworks, distributed services, system kernels, multimedia processing, and artificial intelligence. 
When developers implement business logic by integrating domain-specific development kits (Kits), the actual implementation code of SDK APIs is encapsulated in type declaration files (.d.ets/.d.ts), presenting black-box characteristics to developers. 
This design prevents program static analysis from constructing call graph models for API internal method bodies through conventional interprocedural analysis, and also hinders the establishment of precise cross-layer invocation models. 
However, as the core carrier for system-level capability invocation, if SDK APIs are ignored during pointer analysis, pointer propagation chains will break at system boundaries, leading to premature termination of analysis paths. 
Therefore, \textit{SDKPlugin} is designed to maintain analysis path continuity through predefined pointer propagation rules with specialized object modeling mechanisms.

When \textit{SDKPlugin} detects SDK API interfaces as call targets, it executes return value type verification: when interfaces return types are non-constant references (including class instances, interface types, and other objects with pointer propagation capabilities), these return values serve as potential starting points for subsequent pointer propagation paths.
For this purpose, \textit{SDKPlugin} instantiates abstract heap object nodes matching declared types as logical return value proxies for API calls. 
Although this modeling strategy may introduce abstraction deviations between return values and runtime actual objects, it effectively maintains the integrity of pointer flow propagation chains across system boundaries, ensuring that pointer outflow edges at SDK call points remain valid connections.

\subsubsection{Function Plugin}

In TypeScript, \texttt{Function} is a built-in global type representing the set of all functions. 
It can be considered the "top type" for functions, analogous to how any serves as the top type for all other types. Consequently, any function value can be assigned to a variable of type Function. 
This type provides three core methods—apply, call, and bind, which are instrumental for handling execution contexts (\texttt{this}), dynamic invocation, and metaprogramming by allowing for the explicit specification of \texttt{this} context and arguments. 
While this design offers significant flexibility, it introduces substantial challenges for pointer analysis. 
Therefore, the central task of the \textit{FunctionPlugin} is to accurately model the behavior of these methods and correctly reflect their effects on the PAG.

\begin{figure}
    \centering
    \includegraphics[width=1\linewidth]{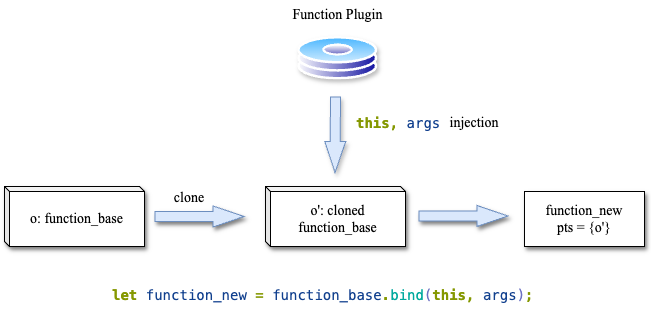}
    \caption{Context and param injection via Function Plugin}
    \label{fig:function_plugin}
    \vspace{-1.4em}
\end{figure}

APAK's standard approach abstracts variables pointing to functions into a specific PAG node type, storing the function as an object model in the node's points-to set. 
The \texttt{Function} type is handled similarly in \textit{FunctionPlugin}. However, the specification of this context and arguments via its methods acts as a form of "secondary processing" on the original function. 
To address this, the \textit{FunctionPlugin} clones the original function's object model and injects the new contextual information into the clone(Figure~\ref{fig:function_plugin}.
This strategy enables customized behavior without altering the original function node. This new, context-aware function node is then immediately invoked for call and apply, or propagated to a new variable in the case of bind for subsequent invocation.

\subsection{Extensibility for Context Sensitivity}
\label{sec:context_sensitive}

During the dynamic execution of a program, the same program point may exist in different runtime context environments~\cite{emami1994context,Whaley2004,Li2018Precision-guided}. These contextual differences lead to multi-version evolution of data flow states at that program point. 
To precisely analyze data flow across these varying contextual paths, APAK already implements function-level sensitivity and call-site sensitivity~\cite{pnueli1981two,Might2010k-CFA-paradox}, with object sensitivity~\cite{Milanova2005object_sensitivity} currently under development. 

Furthermore, APAK is designed for the straightforward extension of its context-sensitivity policies. Through the \texttt{ContextManager}, users can define custom strategies for generating context elements on demand. The framework's base interfaces support the implementation of custom contexts with a depth of up to 5, significantly enhancing the overall extensibility of the analysis.



For specific semantic scenarios, \toolname uses selective context-sensitive strategies~\cite{Smaragdakis2014Introspective, Lu2019Precision-preserving, Lu2021Eagle}. Taking the global object \texttt{globalThis} access scenario as an example: As a cross-context shared global namespace carrier, regardless of how the call stack's context evolves, property access operations on \texttt{globalThis} must maintain a consistent view of global state. If context identifiers are forcibly injected at such access points, it would incorrectly split the data flow of the same global property into multiple contextual versions. Therefore, \toolname selectively suppresses context injection at special semantic nodes like global object access and singleton pattern invocations through static scope determination rules. This mechanism ensures the analyzer can both maintain context sensitivity for regular object access and properly handle data flow homogeneity requirements under global scope.


\section{Evaluation}

We conducted comprehensive evaluation of \toolname using our self-built ArkTS application dataset, with experimental design focusing on three core research questions to systematically validate \toolname's capabilities across effectiveness, efficiency, and practical dimensions:

\begin{itemize}[leftmargin=*]
    \item RQ1: How precise is \toolname in call graph construction?
    \item RQ2: How efficient is \toolname in analysis?
    \item RQ3: How practical is \toolname in resolving real-world problems?
\end{itemize}


\begin{figure}
    \centering
    \includegraphics[width=1\linewidth]{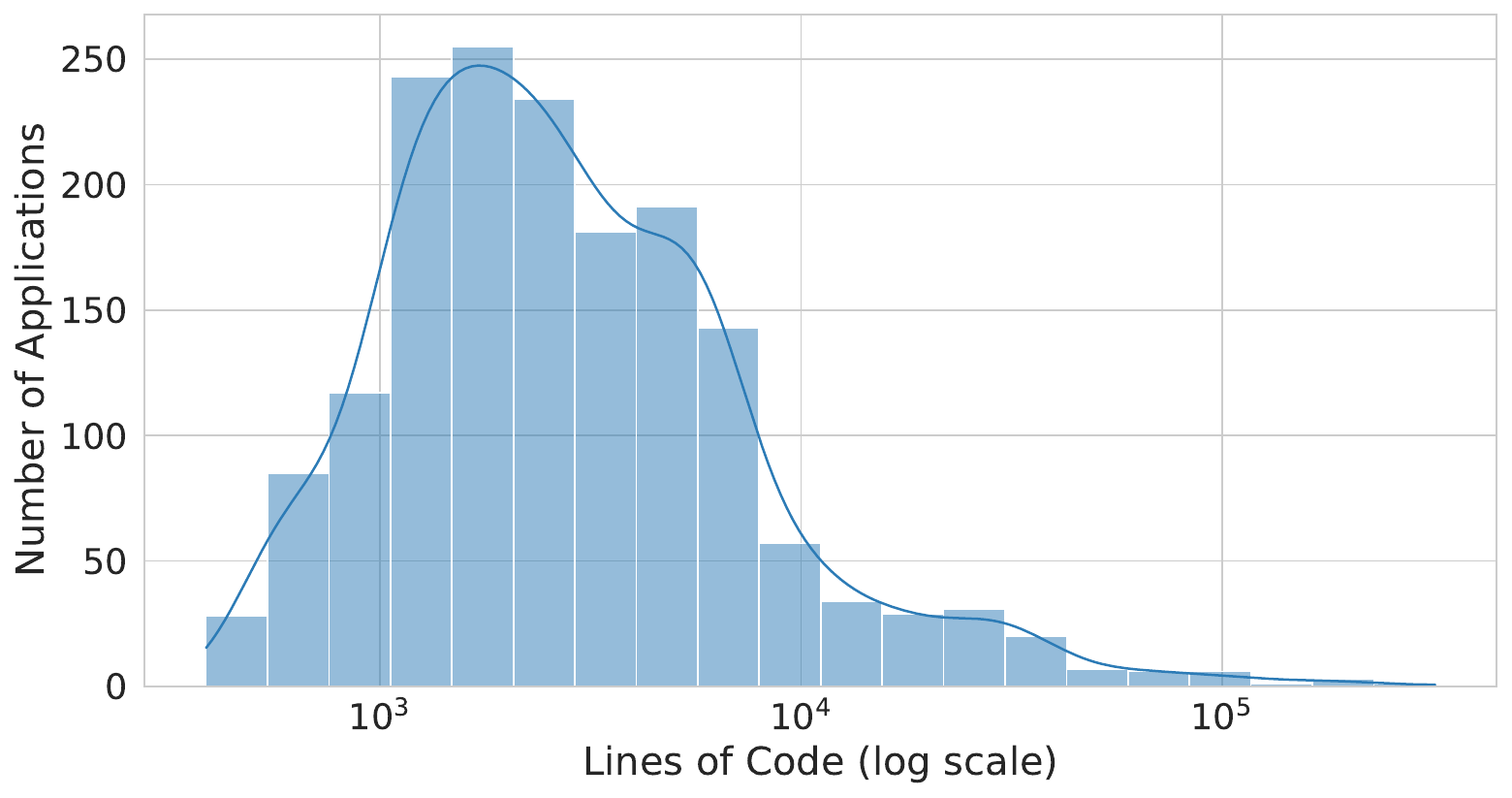}
    \caption{Distribution of collected Apps in our test set.}
    \label{fig:dataset_scale}
    \vspace{-1.4em}
\end{figure}

\noindent \textbf{App Dataset}: The dataset comprises 1,672 production-grade ArkTS applications collected from the OpenHarmony open-source community, covering applications spanning API versions 9 to 12.  It incorporates representative application scenarios including UI component development, cross-device management, and distributed data synchronization, ensuring the evaluation results reflect authentic industrial-scale code characteristics. Its scale is shown in Figure~\ref{fig:dataset_scale}.

\noindent\textbf{Experimental Environment}: The experimental platform configuration is as follows: Ubuntu 20.04.6 LTS, Intel(R) Xeon(R) Gold 6330 CPU @ 2.00GHz, 64GB RAM, Node v23.9.0.

\subsection{RQ1: Accuracy}

We conducted systematic validation on the dataset using a context length of k=2. Through random sampling, 12 application samples (500-7000 LOC) were selected for manual verification to perform quantitative evaluation of the generated call graphs. 
The validation was conducted by two software engineering experts. We first established a comprehensive annotation protocol, complete with precise definitions, rules for handling complex call types, and illustrative examples to ensure consistency. Following this protocol, each expert independently inspected the source code and meticulously labeled every call graph edge. Subsequently, their findings were merged, and any initial disagreements were resolved through a brief joint discussion to reach a final, verified consensus.

The validation was performed by two software engineering experts following a systematic protocol.

Experimental data in Table~\ref{tab:call_graph} demonstrates that \toolname achieves breakthrough precision performance: at the call statement resolution level, its average precision metric consistently approaches the theoretical maximum of 100\%, showing significant false positive reduction compared to the CHA/RTA methods. Simultaneously, \toolname achieves an absolute recall rate improvement exceeding 10\%. This enhancement primarily originates from its precise modeling capability for ArkTS-specific programming paradigms such as dynamic closure binding and asynchronous callback chains, enabling successful capture of call relationships missed by the baseline methods and consequent expansion of coverage scope.

Further experimental evidence confirms that \toolname's precision advantage directly correlates with enhancements in its type constraint system. The implemented inference rule set, combined with ArkTS's semantic constraints, significantly improves object flow resolution accuracy, particularly in handling complex scenarios including polymorphic dispatch and cross-component communication patterns.

\begin{table*}[htbp]
    \centering
    \caption{Comparison between \toolname and existing static analysis methods for the accuracy of call graph construction.}
    \label{tab:call_graph}
    \begin{tabular}{c|c|c|c|c|c|c|}
        
        \multirow{2}{*}{Application} & \multicolumn{2}{c|}{\toolname} & \multicolumn{2}{c|}{CHA} & \multicolumn{2}{c|}{RTA} \\ \cline{2-7}
         & precision & recall & precision & recall & precision & recall \\ \cline{1-7}
        AACommandpackage            & \textbf{100.0\%}   & \textbf{78.5\%}    & \textbf{100.0\%}   & 70.8\%    & \textbf{100.0\%}   & 52.3\%    \\ 
        ActsCleanTempFilesRely      & \textbf{100.0\%}   & \textbf{80.7\%}    & 91.2\%    & 74.4\%    & 94.2\%    & 68.1\%    \\ 
        ActsRegisterJsErrorRely     & \textbf{100.0\%}   & \textbf{85.9\%}    & 89.3\%    & 51.0\%    & 93.0\%    & 42.1\%    \\ 
        AdaptiveServiceWidget       & \textbf{100.0\%}   & \textbf{84.9\%}    & 75.2\%    & 75.2\%    & 75.3\%    & 56.5\%    \\ 
        btmanager\_errorcode401     & \textbf{100.0\%}   & \textbf{92.2\%}    & 97.6\%    & 78.4\%    & 97.4\%    & 72.5\%    \\ 
        CustomCommonEventRely       & \textbf{100.0\%}   & \textbf{94.3\%}    & 94.3\%    & 89.2\%    & 95.9\%    & 78.0\%    \\ 
        formsupplyapplicationC      & \textbf{100.0\%}   & \textbf{92.2\%}    & 94.3\%    & 52.5\%    & 97.1\%    & 43.2\%    \\ 
        NdkOpenGL                   & \textbf{100.0\%}   & 63.3\%    & 93.5\%    & \textbf{73.5\%}    & 91.5\%    & 47.8\%    \\ 
        PixelConversion             & \textbf{100.0\%}   & \textbf{75.9\%}    & 95.4\%    & 52.3\%    & 93.9\%    & 39.2\%    \\ 
        rpcserver                   & \textbf{98.7\%}    & \textbf{79.2\%}    & 96.1\%    & 51.6\%    & 95.8\%    & 47.4\%    \\ 
        server                      & \textbf{100.0\%}   & \textbf{93.6\%}    & 97.1\%    & 43.0\%    & 97.5\%    & 43.8\%    \\ 
        TestExtensionAbility\_001   & \textbf{100.0\%}   & \textbf{86.7\%}    & 97.9\%    & 76.7\%    & 97.1\%    & 56.7\%    \\ 
    \end{tabular}
    \vspace{-2em}
\end{table*}

\begin{figure}
    \centering
    \includegraphics[width=1\linewidth]{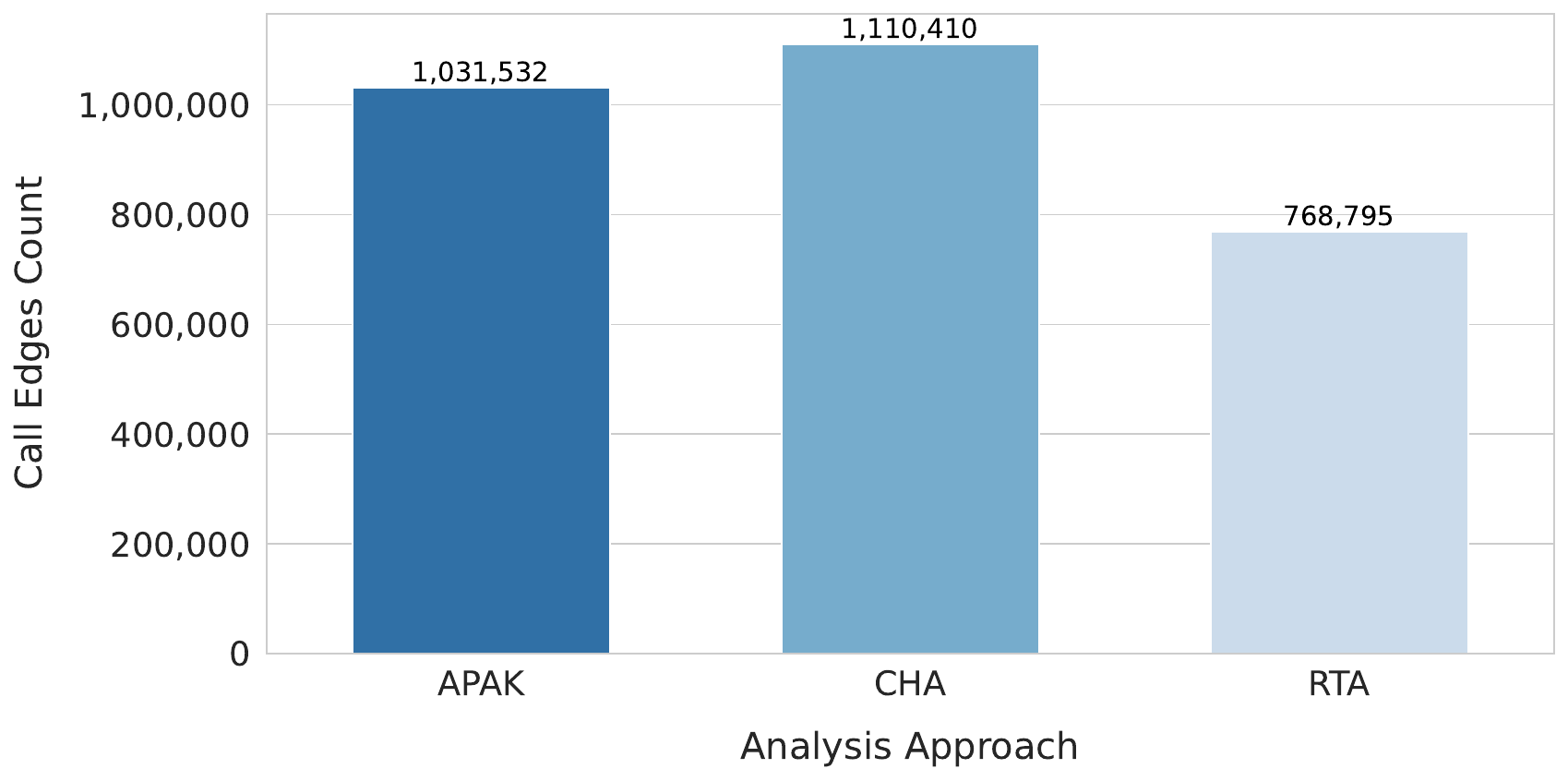}
    \caption{Call edges scale of different approaches.}
    \label{fig:call_edges_scale}
    \vspace{-2em}
\end{figure}

We also conducted full-dataset call edge statistics to compare scale characteristics of call graphs constructed by \toolname, the CHA, and RTA methods (as shown in Figure~\ref{fig:call_edges_scale}). 
Experimental data reveal that the \toolname framework achieves 34.2\% absolute increases in total call edges compared to RTA, demonstrating its superior recall in identifying call relationships that RTA misses. Conversely, compared to the sound but imprecise CHA, \toolname generates a 7.1\% smaller graph. This reduction is a sign of higher precision, as \toolname's powerful pointer analysis effectively prunes a significant number of false-positive edges that CHA incorrectly reports.

\findingbox{}{These empirical evidences quantitatively demonstrate \toolname's completeness advantage in modeling ArkTS language features and validates its effectiveness in call graph construction.}

\subsection{RQ2: Efficiency}

\begin{figure*}[!t]
    \centering
    \begin{subfigure}[b]{0.48\textwidth}
        \includegraphics[width=\textwidth]{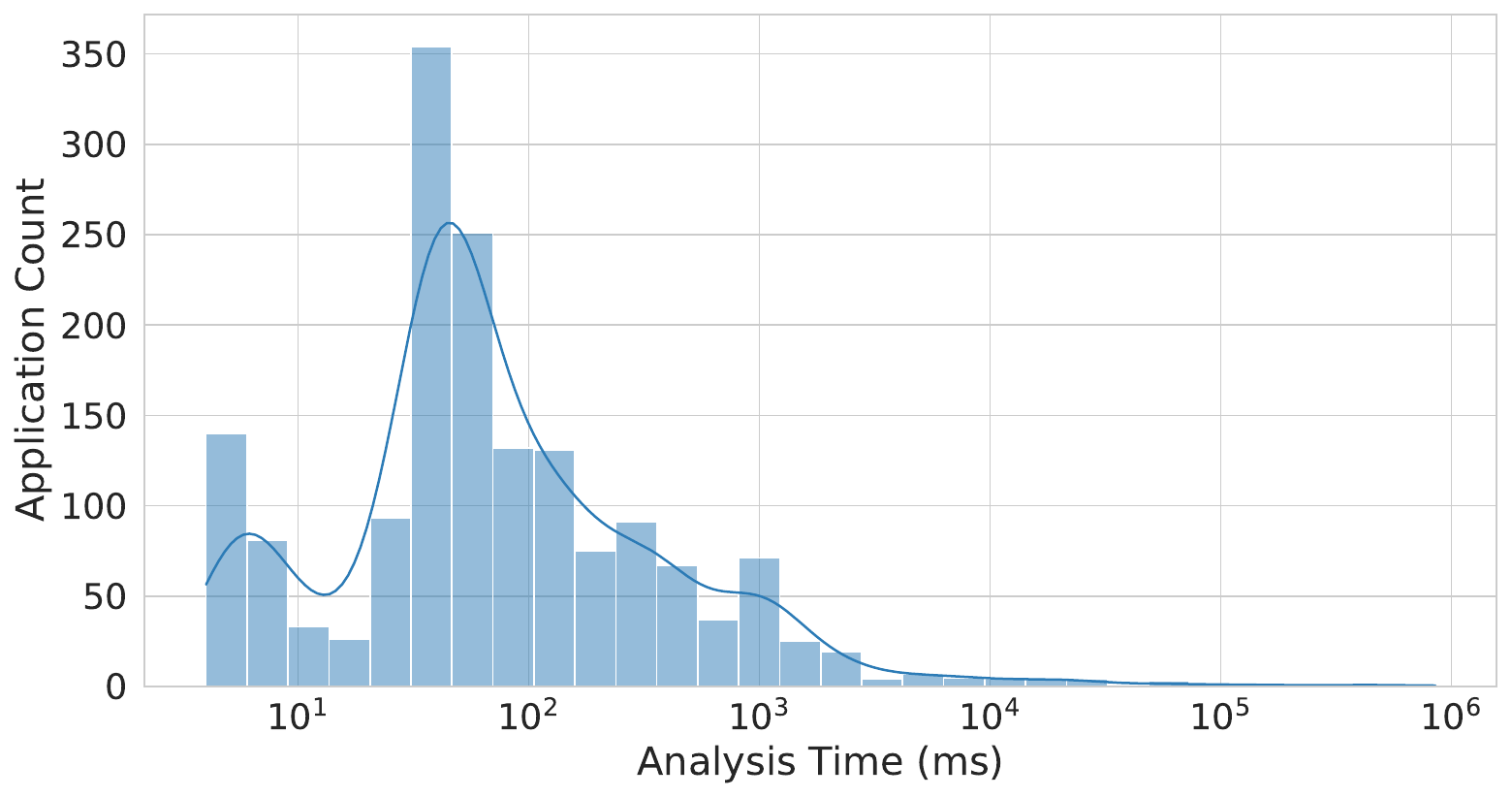}
        \caption{Execution time.}
        \label{fig:time}
    \end{subfigure}
    \hfill
    \begin{subfigure}[b]{0.48\textwidth}
        \includegraphics[width=\textwidth]{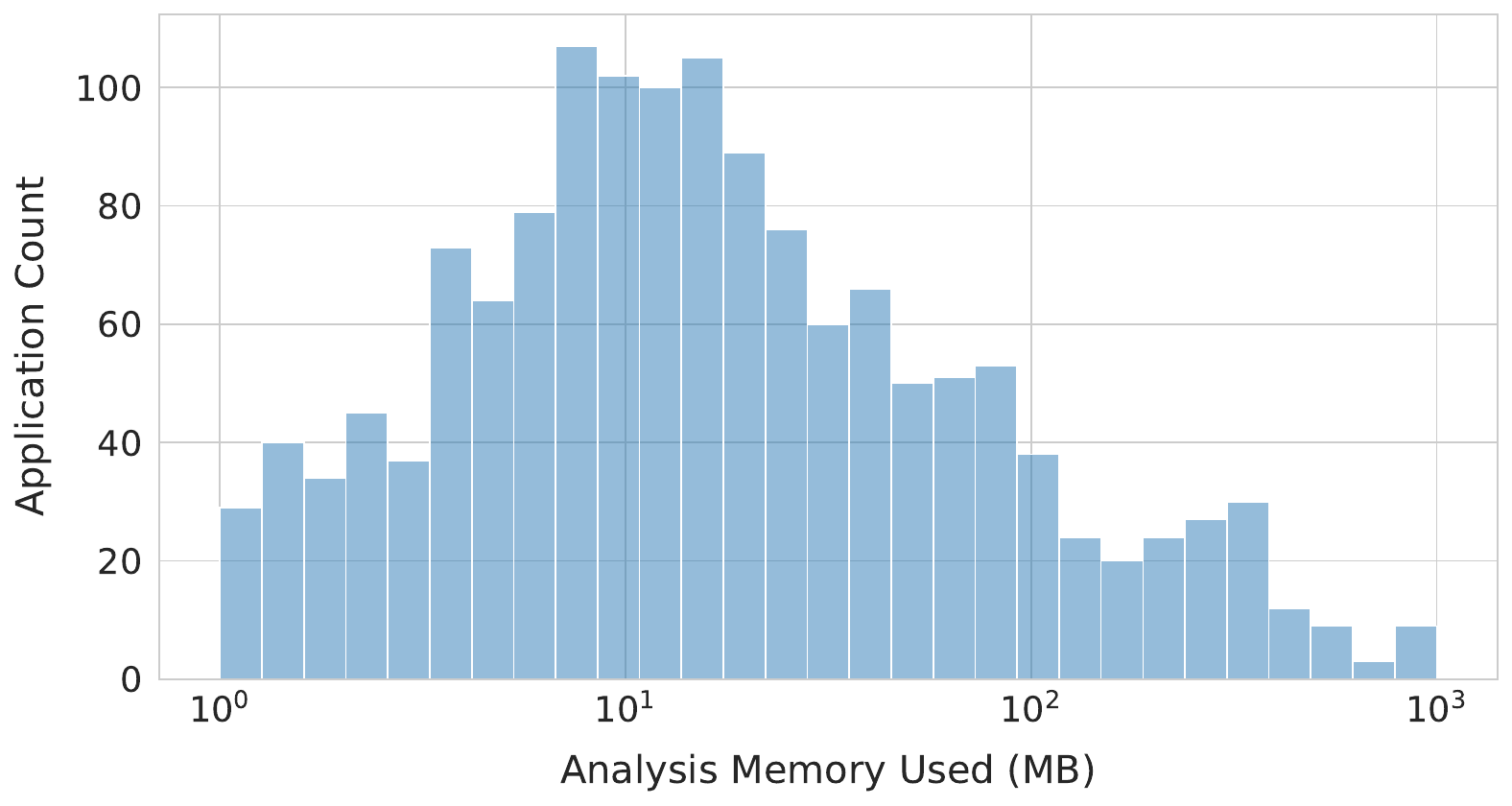}
        \caption{Memory consumption.}
        \label{fig:memory}
    \end{subfigure}
    
    \begin{subfigure}[b]{0.32\textwidth}
        \includegraphics[width=\textwidth]{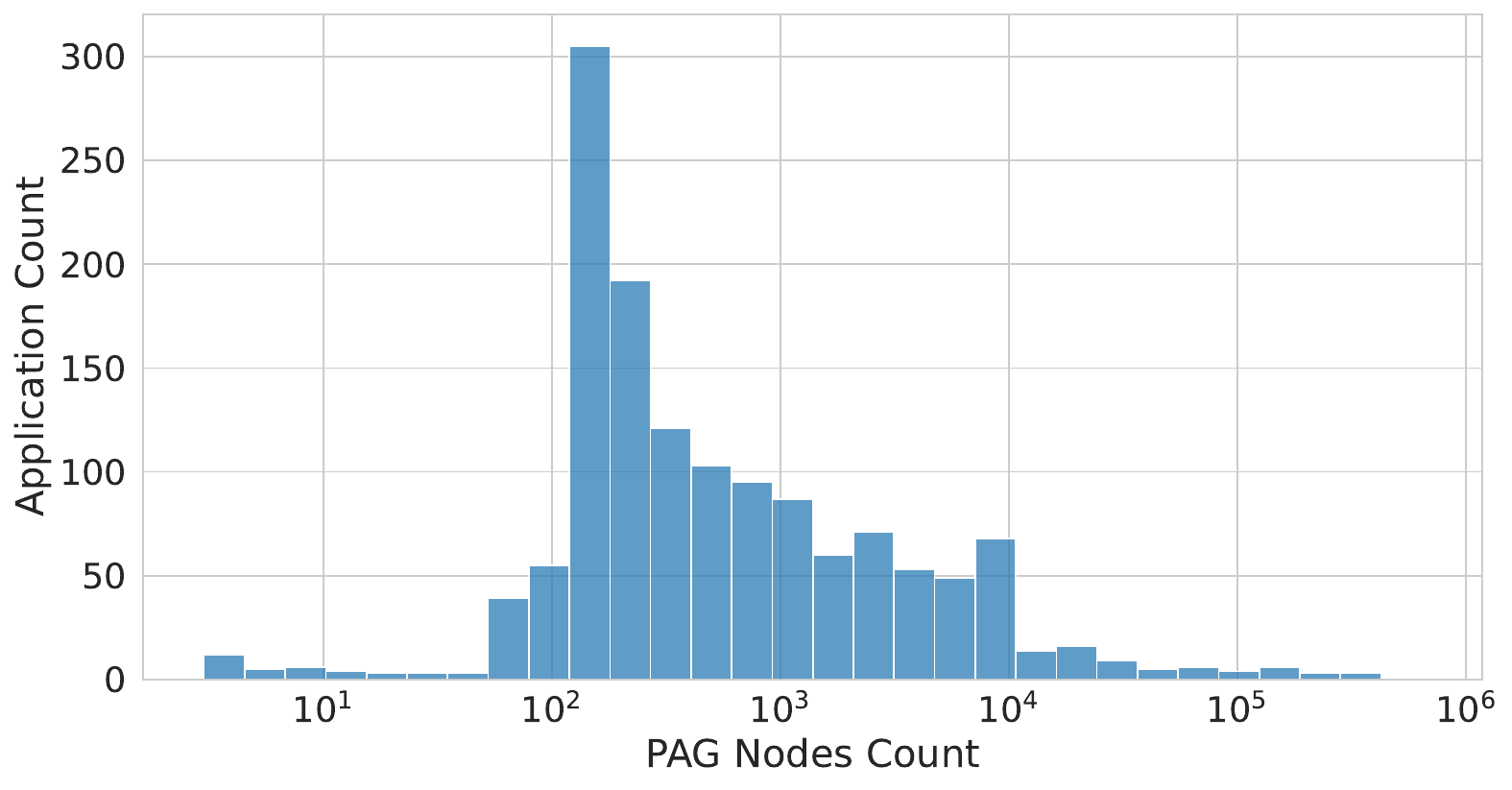}
        \caption{Node scales of PAG.}
        \label{fig:pag_scale}
    \end{subfigure}
    \hfill
    \begin{subfigure}[b]{0.32\textwidth}
        \includegraphics[width=\textwidth]{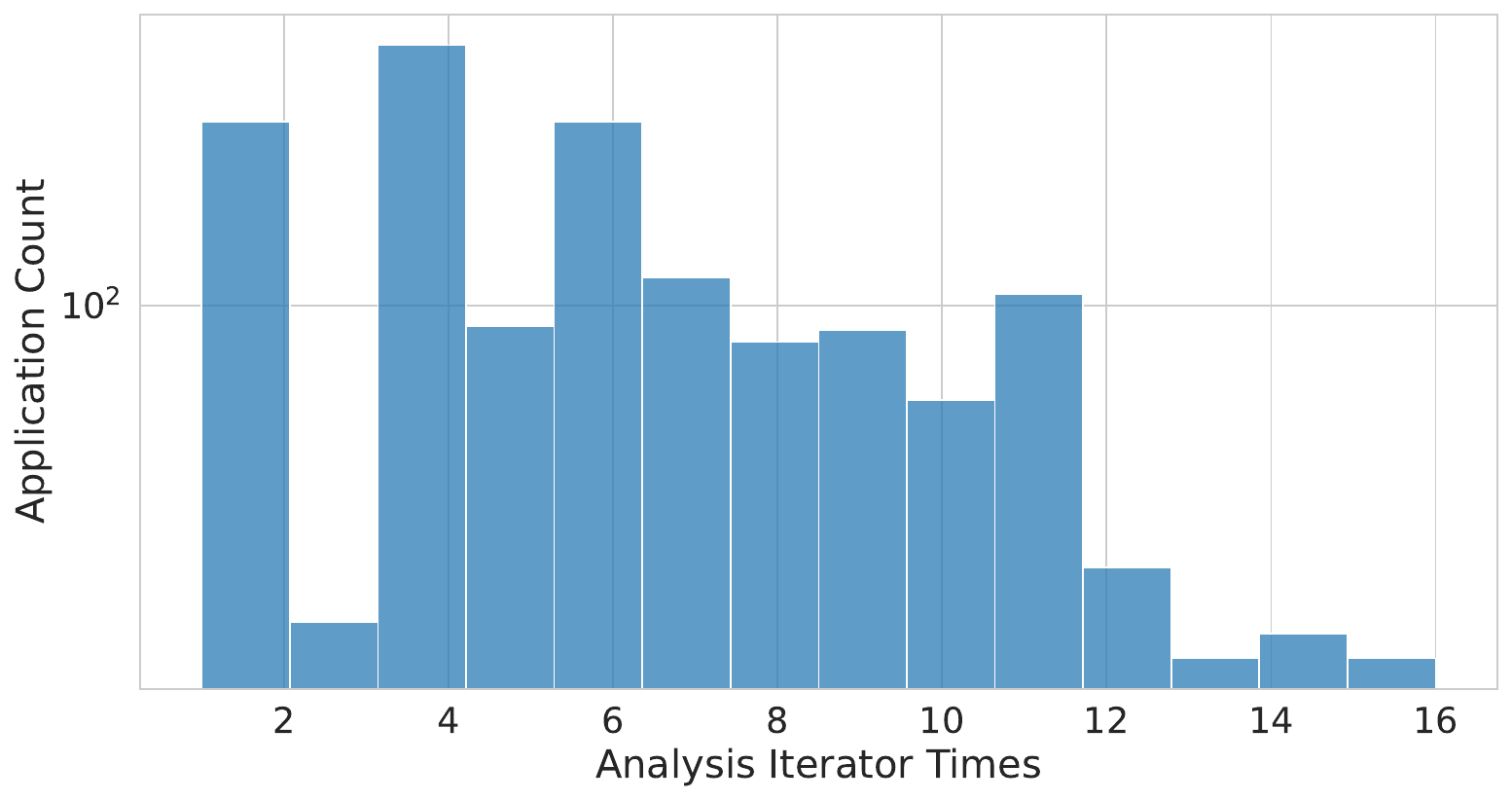}
        \caption{Analysis iterator times.}
        \label{fig:iterator_time}
    \end{subfigure}
    \hfill
    \begin{subfigure}[b]{0.32\textwidth}
        \includegraphics[width=\textwidth]{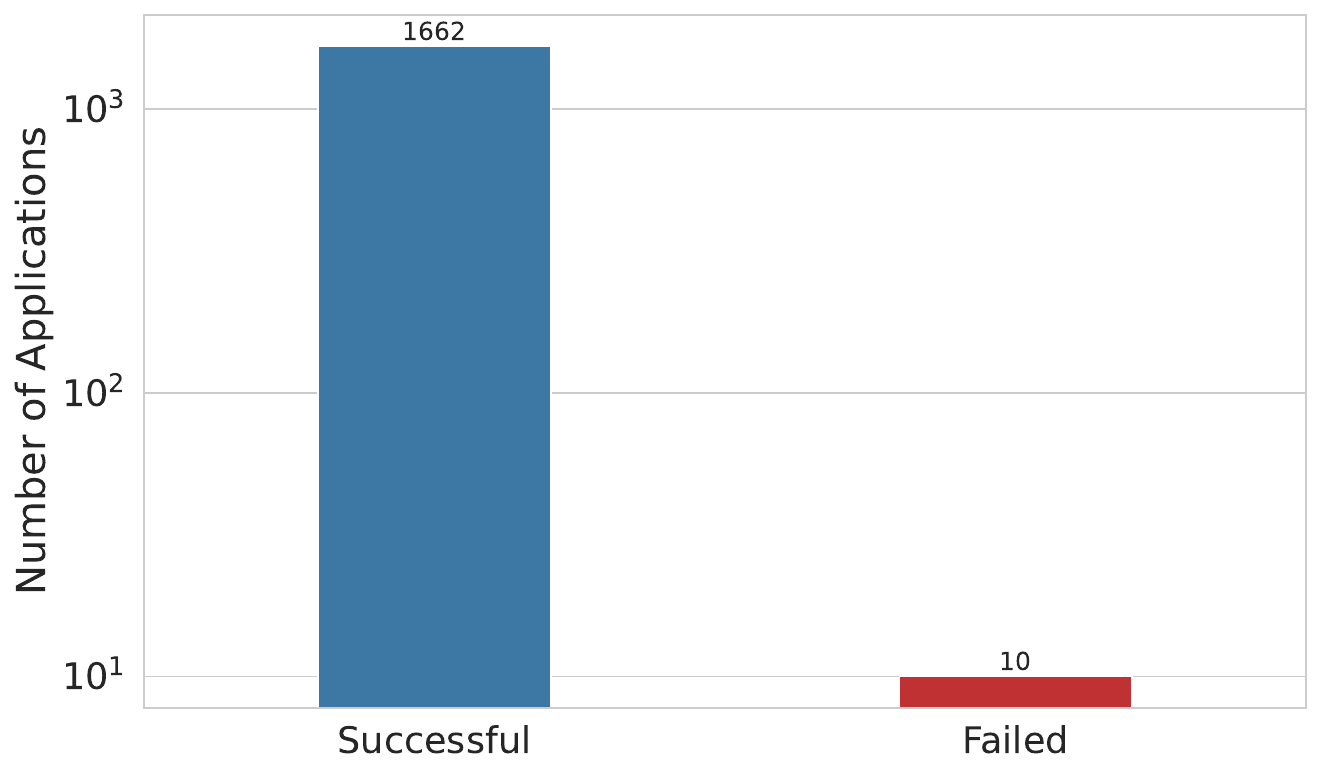}
        \caption{Analysis success rate.}
        \label{fig:analysis_success_rate}
    \end{subfigure}
    
    \caption{Analysis efficiency of \toolname on execution times, memory consumption, node scales of PAG, analysis iterator times, and success rates.}
    \label{fig:efficiency}
    \vspace{-1em}
\end{figure*}

We implemented multi-dimensional monitoring of runtime performance metrics in \toolname through code instrumentation. Figure~\ref{fig:efficiency} presents the core performance parameters measured across the application dataset, including per-instance execution time (\ref{fig:time}), peak memory consumption (\ref{fig:memory}), node scale of PAG (\ref{fig:pag_scale}), analysis iterator times (\ref{fig:iterator_time}, reflect the dynamic call depth) and analysis success rate (\ref{fig:analysis_success_rate}) with k=2 callsite context limit. Within a benchmark suite comprising 1,672 industrial ArkTS applications, the \toolname demonstrates a 99.4\% analysis success rate (1,662/1,672, p $<$ 0.001), confirming its engineering applicability in large-scale codebases.

Experimental results reveal superior runtime characteristics for successfully analyzed instances:
\begin{enumerate}[leftmargin=*]
    \item Median analysis execution time: 53 milliseconds (IQR: 33-160 ms)
    \item 95th percentile peak memory consumption: 284.56 MB
    \item PAG node count distribution exhibits stable patterns with median 225 nodes (IQR: 137-1114)
    \item 95th percentile pointer analysis iterator times: 11
\end{enumerate}

Experimental observations reveal a discrepancy between theoretical expectations and empirical measurements regarding the impact of codebase size on memory consumption, execution time, and PAG scale. The data demonstrates no statistically significant positive correlation between these performance metrics and application scale. Through reverse engineering and call chain analysis, we identify code structure complexity – specifically manifested in call depth and dynamic invocation frequency – as the primary determinant of framework runtime performance.

We conducted an in-depth investigation into the observed bimodal distribution phenomenon of the analysis time metric. Experimental data reveals that in the first peak region with a time threshold of $<$ 10ms, the code complexity of the sample set is relatively low, and there exists a significant proportion of static analysis failure scenarios (including literal expressions and unreachable paths with specific syntactic structures). In the secondary peak region ($<$ 100ms interval), the call graph structures across samples demonstrate concentration (call depth 2-5), validating a certain correlation between control flow complexity and static analysis latency.

Notably, 67.9\% of applications in the test suite exhibit shallow call characteristics (mean call depth: 2.8 ± 0.7 layers). This structural characteristic results in the limited effectiveness of context-sensitive rules within \toolname, particularly reducing the precision in polymorphic method resolution and propagation of heap objects.

Furthermore, of the 10 applications that failed during the analysis, 9 failed due to type errors encountered in ArkIR. The remaining application was excluded because its large scale caused the analysis to exceed our 20-minute timeout threshold.

\findingbox{}{Experimental results validate APAK's efficiency in industrial-scale analysis and demonstrate stable runtime performance. This confirms its viability as an efficient static analysis framework for real-world OpenHarmony applications.}

\subsection{RQ3: Practicality}


In this part, we will show the practicality of our proposed \toolname by 
evaluating our method on a large-scale commercial app (1.2 million lines of code). 



To further validate the efficiency and effectiveness of APAK, we used a complex commercial application with codebase exceeding 1,206,000 lines as a study case, which can help us evaluate the utility of our proposed \toolname in real-world applications. Since we do not have the permission to make this application open-source, we cannot provide any code examples.

Experimental results show that as the callsite context-sensitivity parameter $k$ is incrementally expanded to $k=3$, the single-analysis time remains within the 370-second to 720-second range with gradual increases, while memory consumption exhibits a growing trend (5.4 GB–6.9 GB), and the PAG node scale demonstrates progressive expansion with configurations of $\mathrm{588{,}281}$, $\mathrm{823{,}122}$, and $\mathrm{1{,}067{,}595}$ units respectively. Compared to CHA and RTA, the call edge set size shows substantial increases of 40.0\% and 54.2\% ($k=2$ limit), respectively. As this app is closed-source and has an exceptionally large scale of call graphs (nodes $\geq 8.2 \times 10^5$ and edges $\geq 2.8 \times 10^5$), we cannot manually verify the accuracy between our method and existing static analyzers. However, the results still demonstrate that our method can achieve a more comprehensive analysis than the existing static analysis methods for OpenHarmony.



\findingbox{}{Experimental results demonstrate that APAK exhibits promising feasibility in resolving real-world problems. APAK can effectively address practical challenges in OpenHarmony app data flow analysis, achieving cross-procedural data flow propagation that traditional static type systems cannot handle.}

\section{Threats to Validity}

\paragraph{Internal Threats to Validity} The manual verification process for call edge statistics in RQ1 may introduce human validation bias, particularly in higher-order closure invocation scenarios. The experimental design of RQ2 faces inherent constraints due to the execution environment limitations of commercial closed source systems. The exclusive availability of proprietary applications may lead to potential deviations in performance evaluation metrics.

\paragraph{External Threats to Validity} The pointer analysis technology stack exhibits deep technical dependencies on the completeness of ArkIR. With the continuous evolution of the OpenHarmony ecosystem and ongoing iterative development of ArkAnalyzer, APAK requires persistent updates to maintain compatibility with rapidly changing technical infrastructures.

\section{Related Work}

In C++ ecosystems, the SVF~\cite{Sui2025SVF, Sui2012valueflowanalysis} framework specifically designed for LLVM-based languages provides comprehensive support for interprocedural pointer analysis with value flow tracking capabilities. Within Java environments, frameworks including SPARK~\cite{spark2003}, DooP~\cite{bravenboer2009strictly, Doop2025}, Qilin~\cite{qilin2022}, and Tai-e~\cite{tan2023tai} establish a complete analytical ecosystem, where DooP further extends its applicability to Android application analysis. For JS/TS analysis, TAJS~\cite{Moller2025TAJS} and Jelly~\cite{Moller2025Jelly} employ pattern matching to optimize call graph construction for JavaScript/TypeScript, achieving improvement in vulnerability detection accuracy compared to conventional approaches. Notably, the ArkAnalyzer~\cite{chen2025arkanalyzer} targeting OpenHarmony ecosystem implements standardized processing of ArkTS IR, whose architecture integrates modules like call graph construction.

\section{Conclusion}

This paper addresses the accuracy issue of static analysis for ArkTS, a new programming language, by proposing \toolname. \toolname is the first highly extensible context-sensitive pointer analysis framework specifically designed for OpenHarmony, which achieves higher precision through dual innovation in heap abstraction and API call resolution. 
Experiments demonstrate that \toolname effectively addresses analysis failures and virtual call resolution challenges stemming from ArkUI declarative syntax while maintaining manageable overhead. Future work will focus on enhancing ArkIR parsing capabilities, improving modeling precision for OpenHarmony distributed architectures, and exploring synergistic optimization mechanisms between pointer analysis and taint analysis. 

\section*{Acknowledgements}

This work was partially supported by the National Key Research and Development Program of China (No.2024YFB4506300) and the National Natural Science Foundation of China (No.62572024, No.62502021).

\bibliographystyle{IEEEtran}
\bibliography{main}

\end{document}